# Portfolio Optimization with 2D Relative-Attentional Gated Transformer

Tae Wan Kim
*School of Computer Science*
*University of Sydney*
NSW, Australia
tkim0934@uni.sydney.edu.au

Matloob Khushi
*School of Computer Science*
*University of Sydney*
NSW, Australia
matloob.khushi@sydney.edu.au

*Abstract*—Portfolio optimization is one of the most attentive fields that have been researched with machine learning approaches. Many researchers attempted to solve this problem using deep reinforcement learning due to its efficient inherence that can handle the property of financial markets. However, most of them can hardly be applicable to real-world trading since they ignore or extremely simplify the realistic constraints of transaction costs. These constraints have a significantly negative impact on portfolio profitability. In our research, a conservative level of transaction fees and slippage are considered for the realistic experiment. To enhance the performance under those constraints, we propose a novel Deterministic Policy Gradient with 2D Relative-attentional Gated Transformer (DPGRGT) model. Applying learnable relative positional embeddings for the time and assets axes, the model better understands the peculiar structure of the financial data in the portfolio optimization domain. Also, gating layers and layer reordering are employed for stable convergence of Transformers in reinforcement learning. In our experiment using U.S. stock market data of 20 years, our model outperformed baseline models and demonstrated its effectiveness.

*Keywords—portfolio optimization, reinforcement learning, deep deterministic policy gradient, transformer, relative positional encoding*

## I. Introduction

Portfolio optimization aims to allocate resources optimally into various financial assets to maximize the return while reducing the risks. Since it was theoretically pioneered by [1], many researchers have attempted to solve this problem using various machine learning approaches. Particularly, reinforcement learning is a type of machine learning suitable for sequential decision making such as online portfolio rebalancing. In reinforcement learning, the agent improves its policy to decide an action by repeatedly trying various actions for the environment and maximizing the expected cumulative reward from the environment. This can be implemented by two elements of an agent, the actor that decides its action and the critic that assesses the value of the action with the estimate of the expected cumulative reward. Deep reinforcement learning, a reinforcement learning that utilizes deep neural networks in its actor and critic, is known to be efficient in handling financial problems. However, most research that adopted it in portfolio optimization showed a lack of consideration of realistic constraints, which affects the performance of the models. Moreover, the data in the portfolio optimization domain has intractable characteristics — continuous action space, partial observability, and high dimensionality. This research proposes the Deterministic Policy Gradient with 2D Relative-attentional Gated Transformer (DPGRGT) model to tackle these issues. In the ablation study, the model illustrated profitability as well as stability and outperformed baseline models.

## II. Related Work

Various machine learning approaches have been examined for portfolio optimization. The majority of them, including recent research [2-4], focused on predicting future prices and built portfolios based on the prediction. However, this two-stage approach can be suboptimal in that minimizing the prediction error could be different from the objective of optimizing portfolios and relevant data could be lost by using the predicted price alone [5]. Moreover, the performance of the approach is highly dependent on the prediction accuracy, of which a high degree can be hardly achieved with financial data. For these reasons, researchers employed reinforcement learning as an alternative approach without predicting future prices.

Several researchers [6, 7] used reinforcement learning with discrete action space, simplifying trading actions to include buying, selling, or holding a single asset. Using a limited number of positions in trading was also used in [8], but it is difficult to generalize this approach to apply to large-scale portfolios since expanding the number of assets results in exponential growth in the action space. To address the continuous action space problem, [9] used a policy-based reinforcement learning framework using deep neural networks as its approximation functions and returning the deterministic continuous action values directly from the policy network.

General reinforcement learning approaches are based on Markov Decision Process, which assumes that the current state depends on the previous state only. However, financial markets are only partially observable [10] from the price and volume at a specific point in time. To address the partial observability of the state, reinforcement learning with recurrent neural networks was proposed in [8, 11, 12] using a time series of observations instead of a single observation to represent a state. However, RNN including LSTM still suffers from long-term dependency problems and shows lower performance with longer data [13]. The attention mechanism [14] is proposed to tackle this problem. Particularly, the advent of the Transformer that uses multi-head attention [15] produced state-of-the-art performance in natural language processing and computer vision, but the financial field has not yet benefited from it. Moreover, in [16, 17] Transformer failed to solve a simple Markov Decision Process problem or was comparable to a random model, which suggests that it is extremely difficult to optimize Transformer in a reinforcement learning setting.

There have been many studies on portfolio optimization but most of them are based on unrealistic assumptions about transaction fees and slippage, which have a significant impact on portfolio profitability [18]. In light of this, ignoring or extremely simplifying these constraints makes it difficult to apply the algorithms to real-world asset trading. In this regard, this research proposes a policy-based deep reinforcement learning using a variation of Transformer to address realistic constraints as well as the profitability.

## III. METHODOLOGY

### A. Problem Statement

*1) State*

Since a financial state is only partially observable, this study employs an observation set of historical prices and trading volumes up to time $t$ to represent a state at time $t$. A single observation set $F_t$ is a three-dimensional tensor that consists of five features, Opening, High, Low, Closing prices (OHLC), and trading volumes of assets at time $t$.

$$F_t = (O_t, H_t, L_t, C_t, V_t) \quad (1)$$

Each feature is $t$ by $m + 1$ matrix, where the rows represent the time axis and the columns represent the assets axis that consists of cash and m assets. The opening prices $O_t$, high prices $H_t$, low prices $L_t$, closing prices $C_t$, and trading volumes $V_t$ at time t are as follows:

$$O_t = \begin{pmatrix} o_t^0 & o_t^1 & o_t^2 & \ldots & o_t^m \\ o_{t-1}^0 & o_{t-1}^1 & o_{t-1}^2 & \ldots & o_{t-1}^m \\ o_{t-2}^0 & o_{t-2}^1 & o_{t-2}^2 & \ldots & o_{t-2}^m \\ \vdots & \vdots & \vdots & \ldots & \vdots \\ o_1^0 & o_1^1 & o_1^2 & \ldots & o_1^m \end{pmatrix} \quad (2)$$

$$H_t = \begin{pmatrix} h_t^0 & h_t^1 & h_t^2 & \ldots & h_t^m \\ h_{t-1}^0 & h_{t-1}^1 & h_{t-1}^2 & \ldots & h_{t-1}^m \\ h_{t-2}^0 & h_{t-2}^1 & h_{t-2}^2 & \ldots & h_{t-2}^m \\ \vdots & \vdots & \vdots & \ldots & \vdots \\ h_1^0 & h_1^1 & h_1^2 & \ldots & h_1^m \end{pmatrix} \quad (3)$$

$$L_t = \begin{pmatrix} l_t^0 & l_t^1 & l_t^2 & \ldots & l_t^m \\ l_{t-1}^0 & l_{t-1}^1 & l_{t-1}^2 & \ldots & l_{t-1}^m \\ l_{t-2}^0 & l_{t-2}^1 & l_{t-2}^2 & \ldots & l_{t-2}^m \\ \vdots & \vdots & \vdots & \ldots & \vdots \\ l_1^0 & l_1^1 & l_1^2 & \ldots & l_1^m \end{pmatrix} \quad (4)$$

$$C_t = \begin{pmatrix} c_t^0 & c_t^1 & c_t^2 & \ldots & c_t^m \\ c_{t-1}^0 & c_{t-1}^1 & c_{t-1}^2 & \ldots & c_{t-1}^m \\ c_{t-2}^0 & c_{t-2}^1 & c_{t-2}^2 & \ldots & c_{t-2}^m \\ \vdots & \vdots & \vdots & \ldots & \vdots \\ c_1^0 & c_1^1 & c_1^2 & \ldots & c_1^m \end{pmatrix} \quad (5)$$

$$V_t = \begin{pmatrix} v_t^0 & v_t^1 & v_t^2 & \ldots & v_t^m \\ v_{t-1}^0 & v_{t-1}^1 & v_{t-1}^2 & \ldots & v_{t-1}^m \\ v_{t-2}^0 & v_{t-2}^1 & v_{t-2}^2 & \ldots & v_{t-2}^m \\ \vdots & \vdots & \vdots & \ldots & \vdots \\ v_1^0 & v_1^1 & v_1^2 & \ldots & v_1^m \end{pmatrix} \quad (6)$$

where the subscript of each element stands for the time and the superscript stands for the assets. The elements with the superscript 0 in the first columns stand for cash data and are uniformly set at one.

*2) Action*

The action defined in this research is the proportion of asset investment to be rebalanced at time $t$, since the actions represented in continuous values can be applied to large scale portfolios much better than discrete actions. Action $a_t$ is a portfolio vector at time $t$ where its elements are weights of resource allocation in cash and $m$ assets, and the sum of the weights totals one.

$$a_t = (a_t^0, a_t^1, a_t^2, \ldots, a_t^m), \quad \sum_{i=0}^{m} a_t^i = 1 \quad (7)$$

*3) Reward*

The reward is the risk-adjusted return for the action, with transaction costs applied. The previous portfolio value $p_{t-1}$ is a scalar value calculated by the inner product of the current closing prices of the assets and the previous shares of assets held.

$$p_{t-1} = c_t \cdot s_{t-1} \quad (8)$$

where $c_t$ is a vector of the current closing prices from $c_t^0$ to $c_t^m$ — the first row of the closing price feature $C_t$ — and $s_{t-1}$ is a vector of the previous shares from $s_{t-1}^0$ to $s_{t-1}^m$. The weighted portfolio values $p_{t-1} * a_t$ are modulated to the integer numbers of rebalanced shares $s_t$ of the assets according to the current closing prices as follows:

$$s_t = p_{t-1} * a_t // c_t \quad (9)$$

where // stands for the element-wise floor division operator that returns integer quotients of element-wise division. The transaction fee rate is assumed conservatively at 20 basis points (0.2%), and the slippage rate is set at half of the proportional bid-ask spread. Since the bid-ask spread data is difficult to acquire, the estimate of the proportional bid-ask spread $d_t^i$ [19] is used for a single asset $i$ at time $t$:

$$d_t^i = 2\sqrt{E[(log(c_t^i) - \eta_t^i)(log(c_t^i) - \eta_{t+1}^i)]} \quad (10)$$

where $log(c_t^i)$ is daily closing log-price at time $t$ and $\eta_t^i$ is the average of daily high and low log-prices at time $t$. The rebalancing cost for a single asset is the transaction fee and slippage proportional to the current closing price and change in shares of the asset. Thus, the total rebalancing cost $b_t$ is calculated as follows:

$$b_t = \sum_{i=1}^{m} c_t^i |s_t^i - s_{t-1}^i| (0.2 + 0.5 d_t^i) \quad (11)$$

The rebalanced share for cash $S_t^0$ is the remainder of the previous portfolio value after deduction of the $m$ assets' rebalanced portfolio value and rebalancing cost.

$$s_t^0 = p_{t-1} - \sum_{i=1}^{m} c_t^i s_t^i - b_t \quad (12)$$

Now, the total rebalanced portfolio value $p_t$ can be calculated as follows:

$$p_t = c_t \cdot s_t \quad (13)$$

The return $r_t$ is a log return of the portfolio value.

$$r_t = log(p_t) - log(p_{t-1}) \quad (14)$$

Since the return itself does not reflect risks, the reward function used here is the Sortino ratio [20], which is a variation of the Sharpe ratio [21]. Sharpe ratio is defined as the average of historical returns from $r_1$ to $r_t$ divided by standard deviation of all the returns, whereas Sortino ratio is the expected return divided by the standard deviation of negative returns. The denominators used in both ratios represent the risks of the portfolios.

*B. Model Architecture*

In this section, we propose the Deterministic Policy Gradient with 2D Relative-attentional Gated Transformer (DPGRGT) model. The overall architecture is shown in Fig. 1 and is designed in consideration of the characteristics of the portfolio optimization domain data — continuous action space, partial observability, and high dimensionality. The agent basically follows the structure of Deep Deterministic Policy Gradient [9] for continuous action space, utilizing Transformer encoders whose structure is robust to long-term dependencies of partial observability. Specifically, a variation of Transformer called 2D Relative-attentional Gated Transformer (RG-Transformer) is used as a core part of its actor, target actor, critic, and target critic networks to deal with high dimensional portfolio data.

*1) Deep Deterministic Policy Gradient*

The agent uses a deep neural network as a policy approximator that returns actions with continuous values in a deterministic way [9]. In addition to the actor $\mu$ and critic $Q$ with weights $\theta^\mu$ and $\theta^Q$, respectively, a separate pair of a target actor $\mu'$ and a target critic $Q'$ with respective weights $\theta^{\mu'}$ and $\theta^{Q'}$ is introduced to ensure stable learning. The target return $G_i$ for the $i$-th sample from replay buffer is as follows:

$$G_i = r_i + \gamma\, Q'\left(s_i', \mu'\left(s_i' \middle| \theta^{\mu'}\right) \middle| \theta^{Q'}\right) \quad (15)$$

where $s_i$, $a_i$, $r_i$, $s_i'$ and $\gamma$ present the state, action, reward, next state, and discount factor, respectively. The critic weights $\theta^Q$ are updated by minimizing the loss $L$ from temporal difference error between $G_i$ and $Q(s_i, a_i|\theta^Q)$:

$$L = \tfrac{1}{N}\Sigma_i\bigl(G_i - Q(s_i, a_i|\theta^Q)\bigr)^2 \quad (16)$$

Also, the policy gradient to update the actor weights $\theta^\mu$ is calculated using the chain rule as follows:

$$\nabla_{\theta^\mu} J \approx \frac{1}{N}\sum_i \nabla_{\theta^\mu} Q(s,\mu(s|\theta^\mu)|\theta^Q)|_{s=s_i, a=\mu(s_i)}$$

$$= \tfrac{1}{N}\Sigma_i \nabla_a Q(s,a|\theta^Q)|_{s=s_i, a=\mu(s_i)} \nabla_{\theta^\mu} \mu(s|\theta^\mu)|_{s=s_i} \quad (17)$$

Finally, with a predefined update rate $\tau$, the target actor weights $\theta^{\mu'}$ and target critic weights $\theta^{Q'}$ are updated to $\tau\theta^\mu + (1-\tau)\theta^{\mu'}$ and $\tau\theta^Q + (1-\tau)\theta^{Q'}$, respectively.

During training, the actor includes action space noise of [22] for temporally correlated exploration to avoid local optima as follows:

$$a = \mu(s|\theta^\mu) + \mathrm{d}x_t$$

$$= \mu(s|\theta^\mu) + \theta(\mu - x_{t-1})dt + \sigma\sqrt{dt}N(0,1) \quad (18)$$

where $x_t$ is the noise at time $t$ with fixed parameters $\theta$, $\mu$, and $\sigma$ for the noise generation. To accelerate the training procedure, an asynchronous episodic training method [23] was adopted. The episodic training is processed parallelly by multi-simulators that accumulate each group of experience data $s$, $a$, $r$, and $s$ in the experience replay buffer. To encourage the agent to learn highly-rewarded trajectories more often, HMemory saves an episode only when it renews the highest episodic reward, and those saved data are sampled with probability $\rho$.

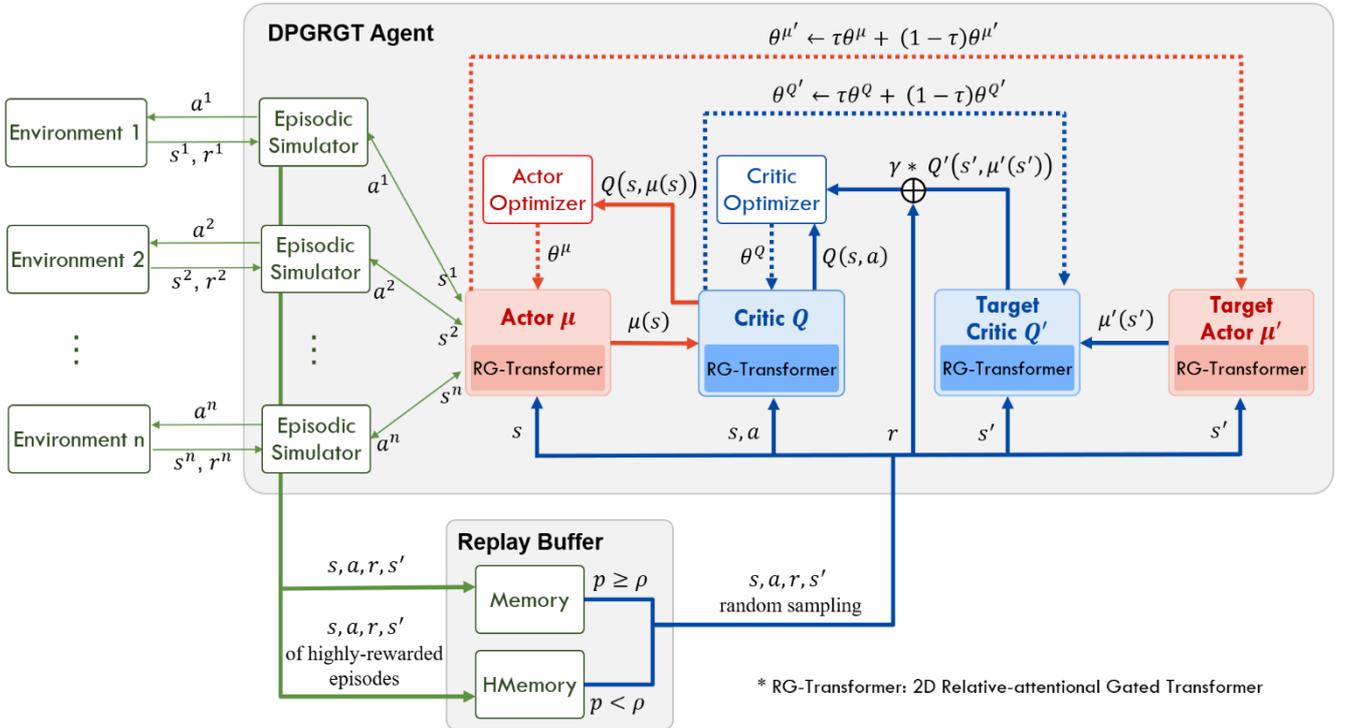

Fig. 1. The overall architecture of DPGRGT model

*2) 2D Relative-attentional Gated Transformer*

The input for portfolio optimization is historical data of partially observable states and is a tensor of three dimensions — financial features, historical time, and assets. To address the high dimensionality, the agent incorporates a variation of Transformer capable of identifying positions of both time and assets into the main part of its actors and critics. Fig. 2 shows the structure of the actor network and its 2D relative positional multi-head attention (2D relative attention), where the length dimension stands for periods while the height dimension represents assets. As seen in Fig. 3, the critic network is similar to the actor network, except that it includes action data as its additional input.

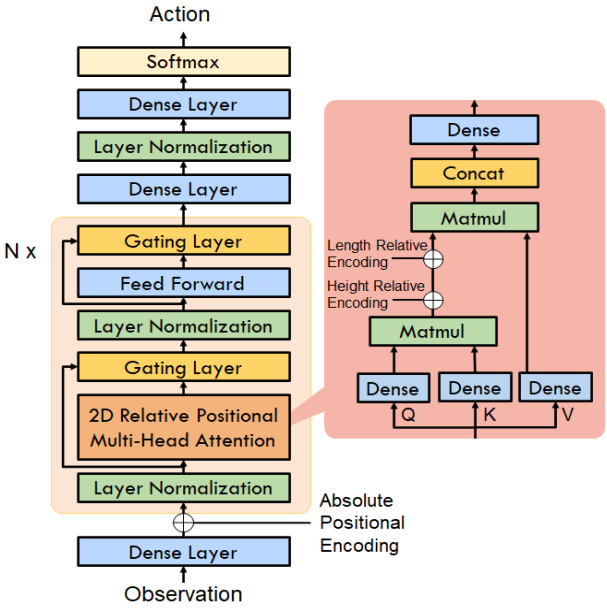

Fig. 2. The actor network and its 2D relative positional multi-head attention of the DPGRGT model

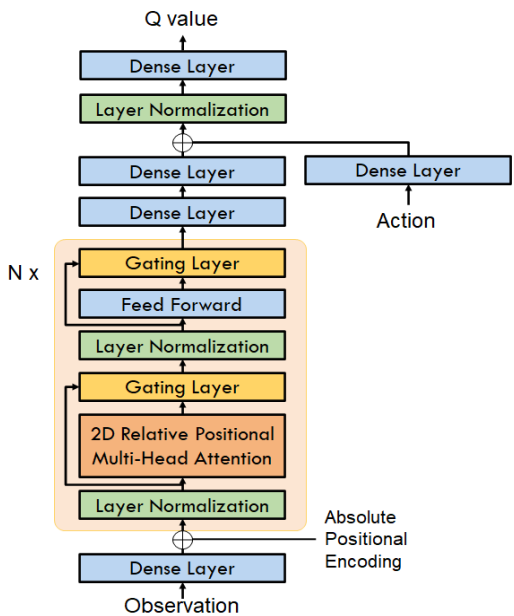

Fig. 3. The critic network of the DPGRGT model

Since the Transformer does not have any recurrent or convolutional structures, it requires additional position information. In addition to the sinusoidal encoding used in the original Transformer, [24, 25] showed the effectiveness of incorporating relative positional information in the self-attention for machine translation and music scoring. For $L \times D$ dimensional data $X$, the relative attention $A_h$ for each head is as follows:

$$A_h = Softmax\left(\frac{XW_h^Q(XW_h^K + R_h)^T}{\sqrt{D_h}}\right) XW_h^V$$

$$= Softmax\left(\frac{XW_h^Q(XW_h^K)^T + XW_h^Q(R_h)^T}{\sqrt{D_h}}\right) XW_h^V \quad (19)$$

where $D_h$ is $D$ divided by the number of heads $h$. $XW_h^Q$, $XW_h^K$, and $XW_h^V$ are an evenly split $L \times D_h$ matrix for each head and are the query, key, and value in the attention, respectively. $R_h$ is a matrix that represents the relative positions between every pair of $X$ elements and is gathered from $L \times D_h$ dimensional relative position embedding $E_h$ learned separately for each head. After multiplying two expanded tensors $XW^Q$ of shape $(h, L, D_h)$ and $(E)^T$ of shape $(h, D_h, L)$, "skewing" [24] the result gives rise to the direct calculation of $XW_h^Q(R_h)^T$ for each head with efficient use of memory. To implement 2D relative attention in our model, two relative position embeddings are used for $L \times H \times D$ dimensional financial data $X'$ where the additional dimension $H$ stands for the height. Embedding $E^l$ and $E^h$ learn the relative positional representation for each pair of data elements in $L$ (time) dimension and $H$ (assets) dimension, respectively. While $X'W^Q$, $X'W^K$, and $X'W^V$ are flattened into $(h, L \cdot H, D_h)$ tensors for matrix multiplication except for the second term in (19), $X'W^Q$ with the original shape of $(h, L, H, D_h)$ is multiplied by the height embedding $(E^h)^T$ of shape $(h, D_h, H)$ and $X'W^Q(E^h)^T$ is flattened into a $(h \cdot L, H, H)$ tensor for skewing. Similarly, after multiplying permuted $X'W^Q$ of shape $(h, H, L, D_h)$ and $(E^l)^T$ of shape $(h, D_h, L)$, the result is flattened to a $(h \cdot H, L, L)$ tensor for skewing. After skewing, both of the calculation outputs result in a tensor of shape $(h, H \cdot L, H \cdot L)$ and are added to the first term instead of the original second term in (19), representing relative positions of both height and length dimensions of the data.

Lastly, the layer normalization is placed before the multi-head attention, and the gating layer is used to replace the residual connection to enhance the stability of Transformers in reinforcement learning [17]. The gating layer utilizes the structure of GRU [13] cell as a gating function $g$ as follows:

$$r = \sigma(W_r y + U_r x)$$

$$z = \sigma(W_z y + U_z x - b_g)$$

$$h = tanh\left(W_g y + U_g(r \odot x)\right)$$

$$g(x, y) = (1 - z) \odot x + z \odot h \quad (20)$$

where $y$ is the input from the previous layer, $x$ is the residual value, and $\odot$ is element-wise multiplication.

## IV. EXPERIMENTS

### A. Experimental Setup

The model is trained and evaluated with assets of nine Dow Jones companies representing each sector - industrials (MMM), financials (JPM), consumer services (PG), technology (AAPL), health care (UNH), consumer goods (WMT), oil & gas (XOM), basic materials (DD) and telecommunications (VZ). The OHLC prices and trading volume data of the assets are collected from Yahoo Finance. The data for 18 years from 2000 to 2017 are used for training, and the data for the period from 2018 to April 2020 is used for evaluation. Each daily observation set consists of historical data for the recent 50 days, and all the data is log differenced for the stationarity of the time series. The maximum length of a single episode is set at 50 days, and the initial investment at 100,000 USD.

Two separate Adam optimizers are used with mini-batch size 32 to train the actor and critic. The learning rate of the actor and critic is 1e-4, and the update rate $\tau$ for the target actor and target critic is 0.15. Discount factor $\gamma$ is 0.9, HMemory sampling rate $\rho$ is 0.2, and the number of threads used for parallel training is 5. The action space noise parameters $\theta$, $\mu$, and $\sigma$ are 0.13, 0, and 0.2, respectively. For the Transformer, three encoder layers, eight heads, 128-dimensional vectors for attention hidden layers, and 512-dimensional vectors for feed-forward network layers are used. All programs were implemented with Python 3 and TensorFlow 2 using Google Colab.

As baseline models, two traditional portfolio models that implement MPT (Markowitz's Modern Portfolio Theory) and Uniform Constant Rebalanced Portfolio (UCRP) strategy [26] are employed. For the ablation study, simple Deep Deterministic Policy Gradient (DDPG), DDPG with Transformer (DDPG_TF), DDPG with 2D Relative-attentional Transformer (DDPG_RP_TF), and DDPG with Gated Transformer (DDPG_GL_TF) are tested under the same conditions. Both the cumulative return and the annualized Sharpe ratio are evaluated to verify the robustness of the models to risks as well as their performances.

### B. Results

Fig. 4 shows the changes in portfolio values for all the models tested with the asset data for the 28 months prior to the experiment. The steep drop seen around March 2020 comes from the outbreak of COVID-19, which has made a huge negative impact on the portfolio values. The portfolio value of the DPGRGT model keeps increasing until it meets the outbreak, and despite the drop, the model outperformed all the other models, proving its resilience and high profitability. DDPG and DDPG with Transformer are slightly better than MPT but show poor performance. DDPG with 2D Relative-attentional Transformer is relatively better than DDPG and DDPG with Transformer, but still worse than the UCRP baseline model. With the exception of DPGRGT, DDPG with Gated Transformer is the only model that is more profitable and better performing than UCRP, which points to the effectiveness of the gating layer of Transformer in reinforcement learning.

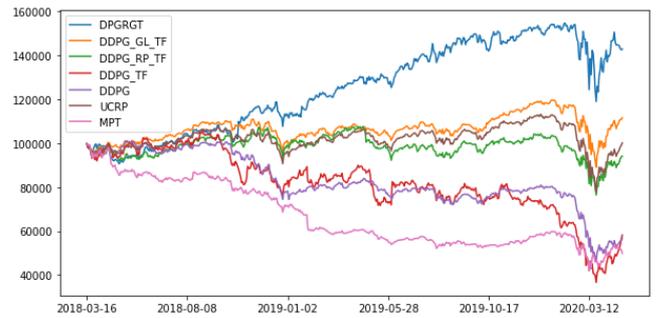

Fig. 4. The portfolio values of the experiment

Table I shows the cumulative returns and the annualized Sharpe ratios of models. While DPGRGT and DDPG with Gated Transformer are profitable, three other models have lost over 40% of their initial value. The MPT model that attempts to maximize its Sharp ratio for every rebalancing chance is the worst, whereas UCRP that rebalances its shares of assets for the constant investment proportion is the second best. This result suggests that transaction fees and slippage have a significant influence on rebalancing performance. With these constraints, the effects of using either 2D Relative-attentional Transformer or Gated Transformer are also limited. Although the pandemic outbreak has undermined the overall performance of the models, DPGRGT that uses both 2D Relative attention and Gated Transformer has ultimately demonstrated stability and strong performance.

TABLE I. THE PERFORMANCE COMPARISON OF MODELS

| Model | Cumulative Return (%) | Annualized Sharpe Ratio |
|---|---|---|
| **DPGRGT (Our Model)** | **43.16** | **0.6418** |
| DDPG_GL_TF | 11.93 | 0.2813 |
| DDPG_RP_TF | -5.45 | -0.1343 |
| DDPG_TF | -41.71 | -0.8191 |
| DDPG | -42.91 | -1.2194 |
| UCRP | 0.53 | 0.0125 |
| MPT | -50.07 | -1.5840 |

## V. CONCLUSION AND FUTURE WORK

This paper proposes a portfolio optimization algorithm based on reinforcement learning using 2D Relative-attentional Gated Transformers. To our best knowledge, this is the first research that applies Transformers to reinforcement learning for portfolio optimization. The experiment shows that the 2D relative attentions and Gating layers improve the performance of Transformers, and combining them creates synergy effects and produces the best results for portfolio optimization. Since even highly profitable models cannot be applied to real trades without stability and realistic constraints taken into consideration, the risk is considered by incorporating the Sortino ratio into the reward function, and the transaction costs are set at a conservative level to ensure the more practical experiment. In a further study, the relations between periods and assets can be analyzed with the multi-head attention weights used in the Transformer to make the model more interpretable and trustable. Furthermore, experiments utilizing a multi-

GPU environment for a wide range of types of assets can be considered for the generalization of the model.